\documentclass{article}

\usepackage{spconf,amsmath,amsfonts,graphicx}
\usepackage{amssymb,textcomp,mathtools}
\usepackage{bm,upgreek,algorithm,hyperref}
\usepackage{multirow,booktabs,hhline,array}
\usepackage{cite,url,makecell,setspace}

\pdfoutput=1

\def\thline{\noalign{\hrule height 1.0pt}}

\renewcommand{\vec}[1]{\bm{\mathrm{#1}}}

\title{Distortion-controlled Training for End-to-end Reverberant Speech Separation with Auxiliary Autoencoding Loss}
\name{Yi~Luo, Cong~Han, Nima~Mesgarani}
\address{Department of Electrical Engineering, Columbia University}

\begin{document}
\ninept
\maketitle

\begin{abstract}
The performance of speech enhancement and separation systems in anechoic environments has been significantly advanced with the recent progress in end-to-end neural network architectures. However, the performance of such systems in reverberant environments is yet to be explored. A core problem in reverberant speech separation is about the training and evaluation metrics. Standard time-domain metrics may introduce unexpected distortions during training and fail to properly evaluate the separation performance due to the presence of the reverberations. In this paper, we first introduce the ``equal-valued contour'' problem in reverberant separation where multiple outputs can lead to the same performance measured by the common metrics. We then investigate how ``better'' outputs with lower target-specific distortions can be selected by auxiliary autoencoding training (A2T). A2T assumes that the separation is done by a linear operation on the mixture signal, and it adds an loss term on the autoencoding of the direct-path target signals to ensure that the distortion introduced on the direct-path signals is controlled during separation. Evaluations on separation signal quality and speech recognition accuracy show that A2T is able to control the distortion on the direct-path signals and improve the recognition accuracy.
\end{abstract}

\section{Introduction}
Recent advances in speech separation have shown superior performance on various datasets \cite{hershey2016deep, isik2016single, yu2017permutation, kolbaek2017multitalker, chen2017deep, luo2017deep, wang2018alternative, luo2019conv, wang2019pitch, liu2019divide, le2019phasebook, xu2019time, luo2020dual, nachmani2020voice, zeghidour2020wavesplit, gu2019end, wang2019speech, wichern2019wham, luo2019fasnet, wang2019sequential, drude2019sms, gu2020enhancing, chen2020continuous, luo2020end, reddy2020interspeech, maciejewski2020whamr}. With the development of end-to-end architectures where the waveforms are used as both the input and the target of the systems, the performance of the state-of-the-art systems significantly outperforms the performance obtained by many oracle time-frequency (T-F) masks which are used as the training target for conventional T-F domain systems. As a second step towards realistic speech enhancement and separation, the community has begun to investigate more challenging datasets, and one of the most important factors in such datasets is the reverberation \cite{wang2018alternative, drude2019sms, wang2019sequential, maciejewski2020whamr, luo2020end}. Reverberation plays an important role in daily communications, and how to make speech enhancement and separation systems work well in reverberant environments remains a very challenging problem.

Both end-to-end systems and T-F domain systems for anechoic speech enhancement and separation can be directly applied in the reverberant scenarios \cite{chen2020continuous}. For systems that do not attempt to perform joint enhancement/separation and dereverberation, a mapping function is typically learned between the reverberant mixture and the reverberant clean signals with either single-channel or multi-channel input \cite{gu2019end, luo2019fasnet, gu2020enhancing, luo2020end}. The evaluation of such systems is also done by comparing the system outputs with the reverberant clean signals. However, using the reverberant clean signal as both the training and evaluation targets introduces new challenges to the current training and evaluation configurations. One core problem, which we refer to as the \textit{equal-valued contour} problem, occurs in many widely-used metrics such as signal-to-noise ratio (SNR) and scale-invariant signal-to-distortion ratio (SI-SDR) \cite{Roux2019sdr}. Equal-valued contour problem denotes the issue that given a reference signal and a metric, there are infinite numbers of estimated signals that can achieve the same performance. Certain estimations among this ``contour'' might be more preferred than the others, however an end-to-end model may lack the ability to distinguish the ``good'' estimations from the ``bad'' ones. As an example of the equal-valued contour problem in reverberant separation, consider an ideal model that always separates the direct-path targets from the reverberant mixture. When evaluated by the signal-level metric between the separation outputs and the reverberant targets, the model will not obtain a high performance especially when the energy of the late reverberation component is large (e.g. with a large reverberation time). However, such an ideal model can achieve very good performance on both word-error-rate (WER) and subjective perceptual quality measures. Suppose there is another model that achieves similar performance as this ideal model when evaluated by the signal-level metric while performs noisy, distorted separation, it is easy to imagine that this model will achieve a much worse performance.

The equal-valued contour problem mainly comes from the training configurations where a single end-to-end training objective is used without further regularizations on the distortion introduced to components such as the direct-path signals. In this paper, we investigate how such regularizations can be incorporated into the training procedure in a simple way. We focus on end-to-end systems where the separation is done by applying a linear mapping, e.g. a multiplicative mask, on the input mixture. This framework includes many recently proposed systems, such as variants of the time-domain audio separation network (TasNet) \cite{luo2019conv, gu2019end, xu2018single, nachmani2020voice, zeghidour2020wavesplit} and any linear neural beamformers \cite{luo2019fasnet, luo2020end}. With the linearity between the input and output, we add an additional auxiliary autoencoding loss term which forces the linear mapping to also perform autoencoding on the direct-path target signal. For example, given a mixture signal which contains one target signal $\vec{x}$ and $K\geq1$ additional interference signals $\{\vec{n}_i\}_{i=1}^{K}$, standard training configuration attempts to optimize the model to learn a linear mapping $\mathcal{M}(\cdot)$ such that $\mathcal{M}(\vec{x} + \sum_{i=1}^K \vec{n}_i) \approx \vec{x}$. The auxiliary autoencoding term corresponds to the reconstruction of the direct-path signal of $\vec{x}$, denoted by $\vec{x}_d$, i.e. $\mathcal{M}(\vec{x}_d) \approx \vec{x}_d$. We refer to this training configuration as the auxiliary autoencoding training (A2T). Auxiliary autoencoding loss has already shown effective in the task of separating varying numbers of sources \cite{luo2020separating}, and here we apply this method in controlling the search space among the equal-valued contours in reverberant separation. We then investigate ways to balance the gradients of the standard objective term and the A2T term for successful training, and conduct experiments with both SNR and SI-SDR training objectives on a simulated noisy reverberant dataset. We also evaluate how do models trained with and without A2T differ when evaluated by WER, and show that A2T is also able to improve the recognition accuracy in various conditions.

The rest of the paper is organized as follows. Section~\ref{sec:problem} describes the problem formulation of end-to-end reverberant speech enhancement and separation. Section~\ref{sec:contour} demonstrates the equal-valued contour problem in details. Section~\ref{sec:obj} introduces the auxiliary autoencoding training framework and methods for gradient balancing. Section~\ref{sec:config} provides the experiment configurations. Section~\ref{sec:result} shows the experiment results and discussions. Section~\ref{sec:conclusion} concludes the paper.

\label{sec:introduction}

\section{End-to-end Reverberant Speech Enhancement and Separation}
\label{sec:problem}
We briefly overview the problem formulation of end-to-end reverberant speech enhancement and separation. Given $M$ channels of inputs and $C$ signal-of-interests (SOIs), the mixture signal at channel $i$ is represented as:
\begin{align}
    \vec{y}_i = \sum_{j=1}^{C} \,\vec{x}_i^{(j)} + \vec{n}_i, \quad i \in 1, \ldots, M
\end{align}
where $\vec{x}_i^{(j)} \in \mathbb{R}^{1\times T}$ is the $j$-th SOI at channel $i$ and $\vec{n}_i \in \mathbb{R}^{1\times T}$ is the interference at channel $i$. In reverberant environment, each SOI is obtained by convolving a clean signal $\vec{c}_i^{(j)} \in \mathbb{R}^{1\times(T-K+1)}$ with a room impulse response (RIR) filter $\vec{h}_i^{(j)} \in \mathbb{R}^{1\times K}$:
\begin{align}
    \vec{x}_i^{(j)} = \vec{c}_i^{(j)} * \vec{h}_i^{(j)}
\end{align}
By decomposing the RIR filter $\vec{h}_i^{(j)}$ into a direct path RIR $\vec{hd}_i^{(j)}$ and a late reverberation RIR $\vec{hr}_i^{(j)}$, the SOI $\vec{x}_i^{(j)}$ can be decomposed into a direct path signal $\vec{x}_{d,i}^{(j)}$ and a late reverberation signal $\vec{x}_{r,i}^{(j)}$:
\begin{align}
\begin{split}
    \vec{x}_i^{(j)} &= \vec{c}_i^{(j)} * \vec{h}_i^{(j)} \\
    &= \vec{c}_i^{(j)} * (\vec{hd}_i^{(j)} +\vec{hr}_i^{(j)} ) \\
    &\triangleq \vec{x}_{d,i}^{(j)} + \vec{x}_{r,i}^{(j)}
\end{split}
\end{align}

The task of speech enhancement or separation is to extract the SOIs given the mixtures. We consider end-to-end systems that generate a linear mapping $T(\cdot)$ between the mixture and each estimated SOI:
\begin{align}
    \left\{\hat{\vec{x}}^{(j)}\right\}_{j=1}^C = T\left(\left\{\vec{y}_i\right\}_{i=1}^M\right)
\label{eqn:linear}
\end{align}

In typical configurations, the training objective is to minimize the discrepancy between the estimated and the target SOIs at a specific reference microphone:
\begin{align}
    \mathcal{L}_{obj} = \sum_{j=1}^{C} \, D\left(\hat{\vec{x}}^{(j)}, \vec{x}_1^{(j)}\right)
\label{eqn:obj}
\end{align}
where $D(\cdot)$ is a metric on the two signals, and we assume that the first channel is selected as the reference channel without loss of generality.

\section{Equal-valued Contours in Common Metrics}
\label{sec:contour}
\begin{figure}[!ht]
	\small
	\centering
	\includegraphics[width=\columnwidth]{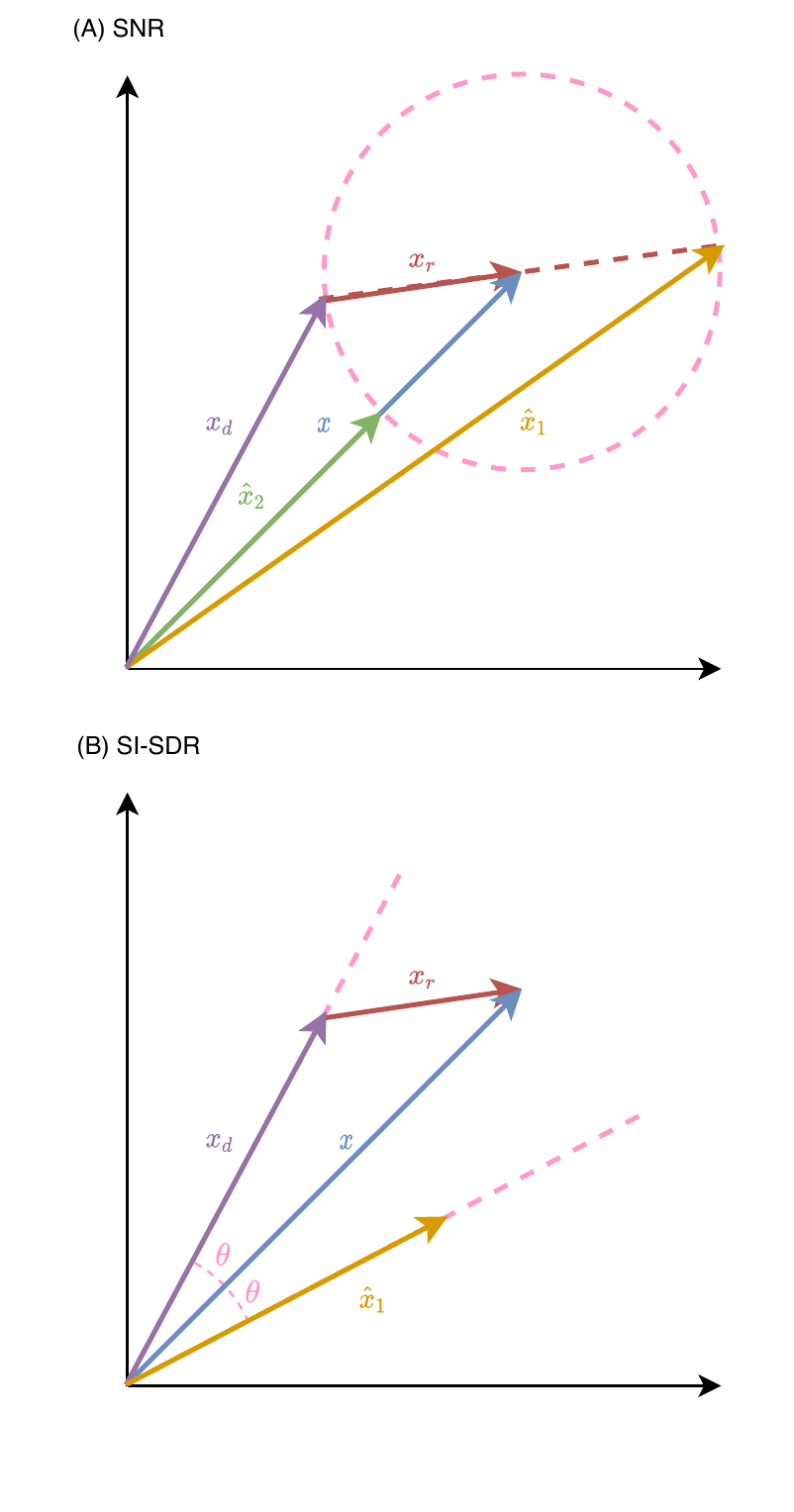}
	\caption{Simplified illustrations for equal-valued contours in (A) SNR metric, and (B) SI-SDR metric.}
	\label{fig:illustration}
\end{figure}

Following the problem formulation in the previous section, we now demonstrate the equal-valued contour problem in two widely-used metrics, namely the SNR and the SI-SDR. SNR between the estimated and target SOIs is defined as:
\begin{align}
    \text{SNR}\left(\hat{\vec{x}}, \vec{x}\right) &= 10\,\text{log}_{10} \frac{||\vec{x}||_2^2}{||\vec{x} - \hat{\vec{x}} ||_2^2} \\
    &= 10\,\text{log}_{10} ||\vec{x}||_2^2 - 10\,\text{log}_{10} ||\vec{x} - \hat{\vec{x}} ||_2^2
\label{eqn:snr}
\end{align}
where the notations are the same as equation~\ref{eqn:obj} despite that we omit the subscripts and superscripts for the sake of simplicity. SNR metric is equivalent to a logarithm-mean square error (log-MSE) metric on the distance between the estimated and target SOIs, thus its equal-valued contours can be defined by the surface of hyperballs whose centers are determined by $\vec{x}$. Figure~\ref{fig:illustration} (A) shows a simplified example of an equal-valued contour in two-dimensional space. The radius of the equal-valued contour in the figure is defined by the reverberation component $\vec{x}_r$, and it's easy to see that $\hat{\vec{x}}_1$, $\hat{\vec{x}}_2$ and $\vec{x}_d$ are on the same contour and have the same SNR value with respect to the reverberant target $\vec{x}$. Moreover, $\hat{\vec{x}}_1=\vec{x}_d+2\vec{x}_r$ adds an additional reverberation component, $\hat{\vec{x}}_2$ is a rescaled version of $\vec{x}$, and $\vec{x}_d$ is the direct-path target. It's obvious that $\vec{x}_d$ is preferred than $\hat{\vec{x}}_2$ and $\hat{\vec{x}}_2$ is preferred than $\hat{\vec{x}}_1$, even though they share a same SNR value.

Another widely-used metric, the SI-SDR, is defined as:
\begin{align}
    \text{SI-SDR}\left(\hat{\vec{x}}, \vec{x}\right) = 10\,\text{log}_{10} \frac{||\alpha\vec{x}||_2^2}{||\hat{\vec{x}} - \alpha\vec{x}||_2^2}
\end{align}
where $\alpha = \hat{\vec{x}}\vec{x}^\top / \vec{x}\vec{x}^\top$ corresponds to the optimal rescaling factor towards the estimated signal. It has been shown in \cite{luo2020separating} that the definition can be rewritten as:
\begin{align}
\begin{split}
    \text{SI-SDR}(\hat{\vec{x}}, \vec{x}) &= 10\,\text{log}_{10}\left(\frac{c(\vec{x}, \hat{\vec{x}})^2}{1-c(\vec{x}, \hat{\vec{x}})^2}\right)
\label{eqn:SI-SDR}
\end{split}
\end{align}
where $c(\vec{x}, \hat{\vec{x}}) \triangleq b/\sqrt{ac} = \hat{\vec{x}}\vec{x}^\top/\sqrt{(\vec{x}\vec{x}^\top)(\hat{\vec{x}}\hat{\vec{x}}^\top)}$ is the cosine similarity between $\vec{x}$ and $\hat{\vec{x}}$. SI-SDR is thus equivalent to the angular distance between the estimated and target SOIs, and its equal-valued contours can be defined by the boundary of cones whose symmetrical axes are defined by $\vec{x}$. Figure~\ref{fig:illustration} (B) shows an example of an equal-valued contour with angle $\theta>0$. Similarly, we have $\hat{\vec{x}}_1$ and $\vec{x}_d$ that share the same value of SI-SDR, while $\vec{x}_d$ is always preferred than $\hat{\vec{x}}_1$.

Note that the definition of equal-valued contours in other metrics, e.g. L1-norm or MSE, can be easily defined in the same way by decoupling the direct-path and reverberation components in the SOIs.

\section{Distortion-controlled Training with Auxiliary Autoencoding Loss}
\label{sec:obj}
\subsection{Auxiliary autoencoding Training}
\label{sec:A2T}

\subsubsection{Definition of A2T}

Auxiliary autoencoding Training (A2T) adds one objective term to control the system outputs on the equal-valued contours. We take $\vec{x}^{(1)}$ as the SOI and omit the subscripts for channel indices where there is no ambiguity. Under the linearity assumption of $T(\cdot)$ in equation~\ref{eqn:linear}, we first rewrite the equation as:
\begin{align}
\begin{split}
    \hat{\vec{x}}_1 &= T\left(\vec{y}\right) \\
    &= T\left(\vec{x}^{(1)} + \sum_{j=2}^C \vec{x}^{(j)} + \vec{n}\right) \\
    &= T\left(\vec{x}^{(1)}\right) + T\left(\sum_{j=2}^C \vec{x}^{(j)}\right) + T\left(\vec{n}\right)
\end{split}
\end{align}
where the system output consist of three parts generated from the direct path, the late reverberation, and the interference, respectively. The conventional training objective sets the reverberant SOI $\vec{x}^{(1)}$ as the training target, and equation~\ref{eqn:obj} becomes:
\begin{align}
    \mathcal{L}_{obj} = D\left(T\left(\vec{x}^{(1)}\right) + T\left(\sum_{j=2}^C \vec{x}^{(j)}\right) + T\left(\vec{n}\right), \vec{x}^{(1)} \right)
\label{eqn:obj-full}
\end{align}

A2T adds an auxiliary autoencoding term on the direct-path signal to the objective:
\begin{align}
    \mathcal{L}_{A2T} &= \underbrace{D\left(T\left(\vec{x}^{(1)}\right) + T\left(\sum_{j=2}^C \vec{x}^{(j)}\right) + T\left(\vec{n}\right), \vec{x}^{(1)} \right)}_{separation} \\
    &+ \underbrace{D\left(T\left(\vec{x}^{(1)}_d\right), \vec{x}^{(1)}_d \right)}_{preservation}
\label{eqn:A2T}
\end{align}
where the auxiliary autoencoding term controls the distortion introduces to the direct-path signal and preserves its signal quality.

To apply permutation invariant training (PIT) in A2T, the output permutations of the two objective terms need to be aligned. In the training phase, PIT is first applied on the separation term to obtain the best label permutation, and the permutation is then applied to the preservation term for auxiliary autoencoding.

\subsubsection{Gradient balancing in A2T}
\label{sec:gradient}
Logarithm-scale objective functions such as SNR and SI-SDR are unbounded and may lead to infinitely large gradients. As autoencoding is a much easier task than separation with a much faster convergence speed, the A2T term may easily dominate the gradients and prevents the standard separation term to be in effect. Thus proper gradient balancing methods need to be applied to ensure successful training. \cite{luo2020separating} proposed the $\alpha$-skewed SI-SDR ($\alpha$-SI-SDR) objective:
\begin{align}
    \text{$\alpha$-SI-SDR}(\vec{x}, \hat{\vec{x}}) \triangleq 10\,\text{log}_{10}\left(\frac{c(\vec{x}, \hat{\vec{x}})^2}{1+\alpha-c(\vec{x}, \hat{\vec{x}})^2}\right)
\label{eqn:a-SI-SDR}
\end{align}
where the gradient scale with respect to the cosine similarity term can be controlled by $\alpha \geq 0$. Similarly, \cite{wisdom2020unsupervised} proposed the $\alpha$-thresholded SNR ($\alpha$-SNR):
\begin{align}
\begin{split}
    \text{$\alpha$-SNR}(\vec{x}, \hat{\vec{x}}) &\triangleq 10\,\text{log}_{10} \frac{||\vec{x}||_2^2}{||\vec{x} - \hat{\vec{x}} ||_2^2 + \alpha ||\vec{x}||_2^2} \\
    &= 10\,\text{log}_{10} ||\vec{x}||_2^2 - 10\,\text{log}_{10} \left(||\vec{x} - \hat{\vec{x}} ||_2^2 + \alpha ||\vec{x}||_2^2\right)
\label{eqn:tau-SNR}
\end{split}
\end{align}
As A2T only serves as a regularization term, we use positive $\alpha$ on the A2T term and set $\alpha=0$ in the separation term. The effect of different values of $\alpha$ as well as the differences between SI-SDR and SNR will be analyzed in Section~\ref{sec:result}.

\subsection{Discussions}
\label{sec:discussion}

\begin{figure}[!ht]
	\small
	\centering
	\includegraphics[width=\columnwidth]{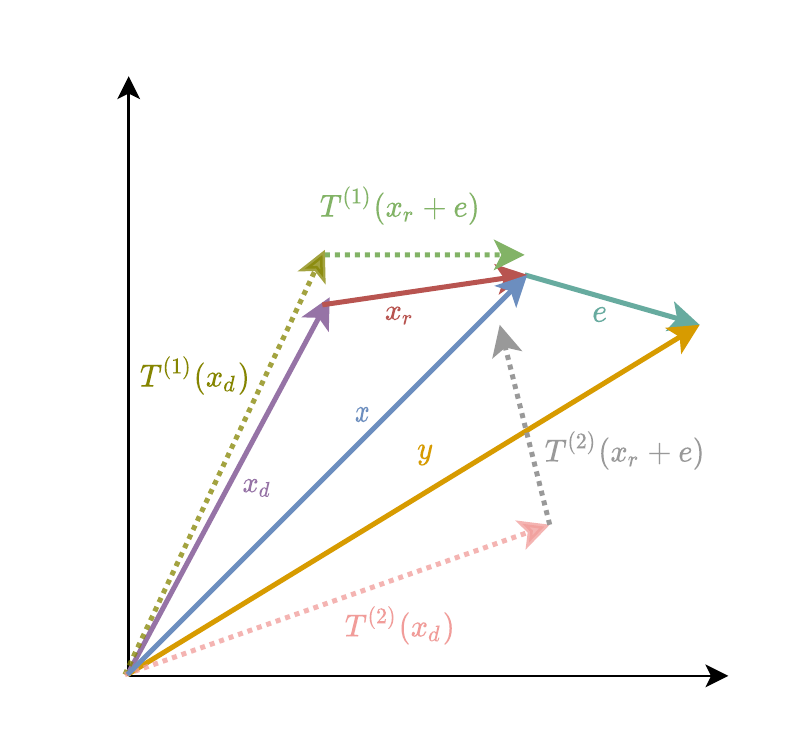}
	\caption{Illustration of two possible linear mappings $T^{(1)}(\cdot)$ and $T^{(2)}(\cdot)$. $T^{(1)}(\cdot)$ denotes the one learned with A2T with a controlled distortion on the direct-path signal $\vec{x}_d$. $T^{(2)}(\cdot)$ corresponds to an unconstrained mapping where the distortion on $\vec{x}_d$ can be significant.}
	\label{fig:A2T}
\end{figure}

The A2T objective term can be directly connected to the optimization target of distortionless response beamformers, such as the MVDR and MPDR beamformers \cite{van2004optimum}, where a distortionless constraint is imposed on the direction of the SOI. A2T does not use such explicit hard constraint, but adds an auxiliary term in the objective as a soft constraint to control the distortion introduced to the direct-path signal. Literatures on imposing constraints on differential frameworks, e.g. the problem of constrained differential optimization \cite{platt1988constrained}, have been investigated in neural networks \cite{pathak2015constrained}, however soft constraints have shown better performance and easier implementation than hard constraints \cite{marquez2017imposing}. Moreover, unlike MVDR/MPDR which are designed only for multi-channel systems, A2T can be easily applied to any end-to-end system which satisfies the linearity assumption of the mapping.

On the other hand, forcing the outputs to meet the A2T constraint may help with the generalization of the model. Figure~\ref{fig:A2T} shows two example mappings $T^{(1)}(\cdot)$ and $T^{(2)}(\cdot)$ with and without A2T, respectively. When the distortion introduced to the direct-path $\vec{x}_d$ is significant, the mapping on the other sources and the noise, i.e. $\vec{e}=\vec{y} - \vec{x}$, needs to compensate for the distortion in order to map to the SOI $\vec{x}$. As the SOI and the interferences are in general uncorrelated, learning such a mapping may hurt the performance and the generalization ability of the system. We would also like to clarify that empirically $T^{(2)}(\cdot)$ might not be the usual case for models without A2T, as in Section~\ref{sec:result} we will show that standard objectives inherently preserve the direct-path signal to some extent. Nevertheless, properly adding the A2T term can almost always achieve on par or better separation performance with a significantly lower distortion on the direct-path signal. This indicates that A2T is able to find ``better'' outputs on the equal-valued contours.

For the consideration of extra computational costs during training, applying autoencoding on the direct-path signal only requires the calculation of the linear transform on the direct-path targets. The complexity for the backward pass is slightly increased as the gradients with respect to the A2T term also need to be backpropagated, but the overall increase on the computational cost is minor.

\section{Experiment configurations}
\label{sec:config}
\begingroup
\setlength{\tabcolsep}{1pt}
\begin{table*}[!ht]
	\small
	\centering
	\caption{Comparison of DPRNN-TasNet models with objectives with and without A2T on the noisy reverberant separation task. ``OR'' stands for the overlap ratio between the two speakers.}
	\label{tab:result-separation}
	\begin{tabular}{c|c|c|c|c|c|c}
		\thline
		\multirow{2}{*}{Objective} & \multirow{2}{*}{$\alpha$} & \multicolumn{5}{c}{SNR / TSNR / SI-SDR / TSI-SDR (dB)} \\
		\cline{3-7}
		& & $\text{OR} \in [0, 25)\%$ & $\text{OR} \in [25, 50)\%$ & $\text{OR} \in [50, 75)\%$ & $\text{OR} \in [75, 100]\%$ & Overall \\
		\thline
		SNR & -- & 14.0 / 18.4 / 13.7 / 18.3 & 10.3 / 12.9 / 9.7 / 12.6 & 8.0 / 10.3 / 6.9 / 9.7 & 6.1 / 8.4 / 4.5 / 7.7 & 9.6 / 12.5 / 8.7 / 12.1 \\
		\hline
		\multirow{8}{*}{+ A2T} & 0 & -0.4 / \textbf{60.6} / -0.4 / \textbf{62.1} & -0.4 / \textbf{60.1} / -0.4 / \textbf{61.9} & -0.4 / \textbf{59.6} / -0.4 / \textbf{61.7} & -0.5 / \textbf{58.8} / -0.5 / \textbf{61.3} & -0.4 / \textbf{59.8} / -0.4 / \textbf{61.8} \\
		& 0.01 & 12.8 / 26.0 / 12.6 / 26.0 & 8.3 / 23.3 / 7.9 / 23.4 & 5.3 / 23.4 / 4.8 / 23.5 & 2.8 / 24.0 / 2.3 / 24.3 & 7.3 / 24.2 / 6.9 / 24.3 \\
		& 0.03 & 13.8 / 23.0 / 13.5 / 23.2 & 9.8 / 18.9 / 9.3 / 19.0 & 7.3 / 16.9 / 6.4 / 16.9 & 5.2 / 15.6 / 3.9 / 15.7 & 9.0 / 18.6 / 8.3 / 18.7 \\
		& 0.1 & 14.0 / 21.4 / 13.7 / 21.5 & 10.1 / 16.8 / 9.5 / 16.8 & 7.8 / 14.6 / 6.9 / 14.5 & 5.9 / 13.1 / 4.5 / 13.0 & 9.5 / 16.5 / 8.7 / 16.5 \\
		& 0.3 & 14.2 / 20.3 / 13.9 / 20.4 & \textbf{10.5} / 15.7 / \textbf{10.0} / 15.6 & \textbf{8.2} / 13.2 / \textbf{7.2} / 13.1 & \textbf{6.2} / 11.5 / \textbf{4.8} / 11.2 & \textbf{9.8} / 15.2 / \textbf{9.0} / 15.1 \\
		& 1 & 14.2 / 19.3 / 13.9 / 19.2 & 10.4 / 14.5 / 9.9 / 14.4 & 8.1 / 12.0 / 7.1 / 11.8 & 6.0 / 10.1 / 4.5 / 9.8 & 9.7 / 14.0 / 8.9 / 13.8 \\
		& 3 & \textbf{14.3} / 18.9 / \textbf{14.0} / 18.9 & \textbf{10.5} / 13.5 / 9.9 / 13.3 & 8.1 / 10.9 / \textbf{7.2} / 10.5 & 6.1 / 9.0 / 4.5 / 8.5 & \textbf{9.8} / 13.1 / 8.9 / 12.8 \\
		& 10 & 14.1 / 18.5 / 13.7 / 18.4 & 10.2 / 13.0 / 9.6 / 12.7 & 7.8 / 10.3 / 6.7 / 9.7 & 5.7 / 8.3 / 4.0 / 7.5 & 9.4 / 12.6 / 8.5 / 12.1 \\
		\thline
		SI-SDR & -- & -- / -- / 13.9 / 16.9 & -- / -- / 9.8 / 10.5 & -- / -- / 6.9 / 7.6 & -- / -- / 4.4 / 5.5 & -- / -- / 8.8 / 10.2 \\
		\hline
		\multirow{8}{*}{+ A2T} & 0 & -- / -- / -0.5 / \textbf{62.2} & -- / -- / -0.4 / \textbf{62.2} & -- / -- / -0.4 / \textbf{62.3} & -- / -- / -0.5 / \textbf{62.5} & -- / -- / -0.4 / \textbf{62.3} \\
		& 0.01 & -- / -- / 12.4 / 27.0 & -- / -- / 7.6 / 24.9 & -- / -- / 4.3 / 26.3 & -- / -- / 1.8 / 28.3 & -- / -- / 6.6 / 26.6 \\
		& 0.03 & -- / -- / 13.2 / 23.5 & -- / -- / 8.9 / 19.7 & -- / -- / 5.9 / 19.0 & -- / -- / 3.4 / 19.8 & -- / -- / 7.9 / 20.5 \\
		& 0.1 & -- / -- / 13.6 / 21.3 & -- / -- / 9.7 / 16.9 & -- / -- / 6.9 / 14.6 & -- / -- / 4.4 / 13.1 & -- / -- / 8.7 / 16.5 \\
		& 0.3 & -- / -- / 14.0 / 20.8 & -- / -- / 9.8 / 16.0 & -- / -- / 7.0 / 13.6 & -- / -- / 4.5 / 11.9 & -- / -- / 8.8 / 15.6 \\
		& 1 & -- / -- / 13.9 / 19.8 & -- / -- / \textbf{9.9} / 14.9 & -- / -- / 7.2 / 12.3 & -- / -- / 4.7 / 10.5 & -- / -- / 8.9 / 14.4 \\
		& 3 & -- / -- / 13.9 / 19.5 & -- / -- / \textbf{9.9} / 14.4 & -- / -- / \textbf{7.3} / 11.9 & -- / -- / \textbf{4.8} / 10.0 & -- / -- / \textbf{9.0} / 14.0 \\
		& 10 & -- / -- / \textbf{14.1} / 19.4 & -- / -- / \textbf{9.9} / 14.0 & -- / -- / \textbf{7.3} / 11.5 & -- / -- / 4.6 / 9.5 & -- / -- / \textbf{9.0} / 13.6 \\
		\thline
	\end{tabular}
\end{table*}
\endgroup

\begin{table*}[!ht]
	\small
	\centering
	\caption{Comparison of WER from models trained with SNR with and without A2T. ``OR'' stands for the overlap ratio between the two speakers.}
	\label{tab:result-recognition}
	\begin{tabular}{c|c|c|c|c|c|c}
		\thline
		\multirow{2}{*}{Objective} & \multirow{2}{*}{$\alpha$} & \multicolumn{5}{c}{WER (\%)} \\
		\cline{3-7}
		& & $\text{OR} \in [0, 25)\%$ & $\text{OR} \in [25, 50)\%$ & $\text{OR} \in [50, 75)\%$ & $\text{OR} \in [75, 100]\%$ & Overall \\
		\thline
		SNR & -- & 15.5 & 20.8 & 38.4 & 60.2 & 34.2 \\
		\hline
		\multirow{3}{*}{+ A2T} & 0.3 & 15.3 & \textbf{17.5} & 36.4 & \textbf{57.7} & \textbf{32.1} \\
	    & 1 & 15.2 & 18.4 & 36.5 & 58.8 & 32.6 \\
	    & 3 & \textbf{14.8} & 20.5 & \textbf{34.5} & 59.5 & 32.8 \\
		\thline
		Noisy reverberant & -- & -- & -- & -- & -- & 17.7 \\
		Clean reverberant & -- & -- & -- & -- & -- & 8.1 \\
		\thline
	\end{tabular}
\end{table*}

\subsection{Dataset}

We evaluate our approach on the task of single-channel noisy reverberant speech separation. The dataset we use is the same as \cite{luo2020end}, which is a simulated two-speaker noisy reverberant dataset. We use the adhoc-geometry configuration which contains 20000, 5000 and 3000 4-second long utterances for training, validation and testing, respectively, and only select the first channel for all experiments. The sample rate for all utterances is 16k Hz. The length and width of the room are randomly sampled between 3 and 10 meters, and the height is randomly sampled between 2.5 and 4 meters. The microphone arrays contain 2-6 microphones with their locations randomly sampled in the room. Two speaker locations are also sampled such that the average speaker angle with respect to the microphone center is uniformly distributed between 0 and 180 degrees. Both microphone and speaker locations are forced to be at least 0.5~m away from the room walls. The noise location is sampled without further constraints.

For the training set, the speech utterances are randomly sampled from the 100-hour Librispeech subset \textit{train-clean-100} \cite{panayotov2015librispeech}. For validation and testing, the speech utterances are sampled from the \textit{dev-clean} and \textit{test-clean} subsets, respectively. A nonspeech noise is randomly sampled from the 100 Nonspeech Corpus \cite{web100nonspeech}. An overlap ratio between the two speakers is uniformly sampled between 0\% and 100\%, and the two speech signals are shifted accordingly and rescaled to a random relative SNR between 0 and 5 dB. The noise is repeated if its length is smaller than 4 seconds, and the relative SNR between the power of the sum of the two clean speech signals and the noise is randomly sampled between 10 and 20 dB. The speech and noise signals are then convolved with a simulated room impulse responses by the image method \cite{allen1979image} using the gpuRIR toolbox \cite{diaz2020gpurir}. The reverberant speech and noise signals are summed to create the mixtures at each microphone.

The direct path room impulse response (RIR) filter in all utterances is defined as the $\pm6$~ms of the first peak in the RIR filter, and the late reverberation RIR filter is defined as the residual of the direct path RIR filter. The direct path and late reverberation signals are obtained by convolution between the speech/noise signals and the corresponding RIR filters.

As the utterances in \cite{luo2020end} are randomly truncated and not suitable for speech recognition evaluation, we simulate another test set with 500 utterances where the utterances are not truncated. In order to approximately match the length of the mixtures in the training set, we randomly sample the utterances from the \textit{test-clean} subset whose length is no longer than 4 seconds. All other configurations are the same as the description above.

\subsection{Model configurations}

We train the single-channel TasNet with objectives with and without A2T regularization. We adopt the DPRNN-TasNet proposed in \cite{luo2020dual} with 128 filters in the encoder/decoder and 6 DPRNN blocks in the separation module. The frame size is always set to 2~ms. All models are trained for 100 epochs with Adam as the optimizer \cite{kingma2014adam}. The initial learning rate is set to 0.001 and is decayed by 0.98 for every two epochs. Gradient clipping of a maximum $L_2$-norm of 5 is applied during training. The training is stopped when no best validation is found for 10 consecutive epochs. No other regularizers or training tricks are applied.

\subsection{Evaluation metrics}

We evaluate the separation performance by both SNR and SI-SDR metrics. To evaluate the distortion introduced to the direct-path signals, we also evaluate the target-SNR (TSDR) and target-SI-SDR (TSI-SDR), which are calculated between the transformed direct-path signal $T(\vec{x}_d)$ and the direct-path signal $\vec{x}_d$:
\begin{align}
    \text{TSNR}\left(\hat{\vec{x}}, \vec{x}\right) &= \text{SNR}\left(T(\vec{x}_d), \vec{x}_d\right) \\
    \text{TSI-SDR}\left(\hat{\vec{x}}, \vec{x}\right) &= \text{SI-SDR}\left(T(\vec{x}_d), \vec{x}_d\right)
\end{align}

We evaluate the speech recognition performance by word error rate (WER). The recognition engine we used is directly taken from the pre-trained transformer model on Librispeech data from ESPNet \cite{watanabe2018espnet}\footnote{https://github.com/espnet/espnet/tree/master/egs/librispeech/asr1}.

\section{Results and discussions}
\label{sec:result}
\begingroup
\setlength{\tabcolsep}{1pt}
\begin{table*}[!ht]
	\small
	\centering
	\caption{Comparison of DPRNN-TasNet models with direct-path RIR filter defined as the $\pm20$~ms of the first peak in the RIR filter.}
	\label{tab:result-separation-20}
	\begin{tabular}{c|c|c|c|c|c|c}
		\thline
		\multirow{2}{*}{Objective} & \multirow{2}{*}{$\alpha$} & \multicolumn{5}{c}{SNR / TSNR / SI-SDR / TSI-SDR (dB)} \\
		\cline{3-7}
		& & $\text{OR} \in [0, 25)\%$ & $\text{OR} \in [25, 50)\%$ & $\text{OR} \in [50, 75)\%$ & $\text{OR} \in [75, 100]\%$ & Overall \\
		\thline
		SNR & -- & 14.0 / 18.7 / 13.7 / 18.6 & 10.3 / 13.1 / 9.7 / 12.8 & \textbf{8.0} / 10.5 / \textbf{6.9} / 10.0 & \textbf{6.1} / 8.6 / \textbf{4.5} / 7.8 & \textbf{9.6} / 12.8 / \textbf{8.7} / 12.3 \\
		\hline
		\multirow{3}{*}{+ A2T ($\pm20$~ms)} & 0.3 & \textbf{14.1} / \textbf{20.3} / \textbf{13.8} / \textbf{20.3} & 10.2 / \textbf{15.3} / 9.7 / \textbf{15.2} & 7.9 / \textbf{12.8} / \textbf{6.9} / \textbf{12.7} & 5.9 / \textbf{10.9} / 4.3 / \textbf{10.7} & \textbf{9.6} / \textbf{14.9} / \textbf{8.7} / \textbf{14.7} \\
		& 1 & \textbf{14.1 }/ 19.4 / \textbf{13.8} / 19.4 & \textbf{10.4} / 14.2 / \textbf{9.8} / 14.0 & 7.9 / 11.5 / \textbf{6.9} / 11.1 & 5.8 / 9.5 / 4.2 / 9.1 & \textbf{9.6} / 13.7 / \textbf{8.7} / 13.4 \\
		& 3 & 14.0 / 19.1 / 13.7 / 19.2 & 10.1 / 13.7 / 9.5 / 13.5 & 7.8 / 11.0 / 6.7 / 10.7 & 5.8 / 9.0 / 4.1 / 8.6 & 9.4 / 13.3 / 8.5 / 13.0 \\
		\thline
	\end{tabular}
\end{table*}
\endgroup

\begin{table*}[!ht]
	\small
	\centering
	\caption{Comparison of WER from models trained with direct-path RIR filter defined as the $\pm20$~ms of the first peak in the RIR filter.}
	\label{tab:result-recognition-20}
	\begin{tabular}{c|c|c|c|c|c|c}
		\thline
		\multirow{2}{*}{Objective} & \multirow{2}{*}{$\alpha$} & \multicolumn{5}{c}{WER (\%)} \\
		\cline{3-7}
		& & $\text{OR} \in [0, 25)\%$ & $\text{OR} \in [25, 50)\%$ & $\text{OR} \in [50, 75)\%$ & $\text{OR} \in [75, 100]\%$ & Overall \\
		\thline
		SNR & -- & 15.5 & 20.8 & 38.4 & 60.2 & 34.2 \\
		\hline
		SNR+A2T ($\pm6$~ms) & 0.3 & 15.3 & \textbf{17.5} & \textbf{36.4} & \textbf{57.7} & \textbf{32.1} \\
		\hline
		\multirow{3}{*}{SNR+A2T ($\pm20$~ms)} & 0.3 & \textbf{14.8} & 19.2 & 38.2 & 62.1 & 34.0 \\
	    & 1 & 15.4 & 19.2 & 36.5 & 61.1 & 33.4 \\
	    & 3 & 15.0 & 19.3 & 36.8 & 60.2 & 33.2 \\
		\thline
	\end{tabular}
\end{table*}

Table~\ref{tab:result-separation} shows the experiment results of models trained with different objective functions, with and without A2T, and with different values of $\alpha$ for gradient balancing in the A2T term. For the models trained with SI-SDR, we do not report the SNR and TSNR scores as SI-SDR does not preserve the scale of the outputs. We first notice that the models trained with original SNR and SI-SDR objectives are able to inherently control the distortion introduced to the direct-path signals to some extent, and in low-overlapped utterances the distortion is significantly lower than high-overlapped utterances. On the one hand, the performance of TSNR and TSI-SDR in low-overlapped utterances is expected as in the nonoverlapped regions the separation model is equivalent to an autoencoding model. On the other hand, the worse performance in high-overlapped utterances shows that the problem mentioned in Section~\ref{sec:discussion} are practical, and the separated outputs might not be the preferred ones among the equal-valued contours. Moreover, SI-SDR objective even leads to a lower TSI-SDR score than the SNR objective across all overlap ratios. Since the SNR objective is also able to preserve the output scale, the results indicate that SNR can be a good replacement of SI-SDR as a training objective even when the evaluation is done by SI-SDR. This also matches the observation in \cite{ochiai2020beam} where SNR led to at least on par performance as SI-SDR.

We then notice that for $\alpha=0$ in the A2T term, models trained with both objectives fail to converge and the gradients are dominated by the A2T term. It leads to a significantly higher performance on the TSNR and TSI-SDR scores but completely fails on separation. The TSNR and TSI-SDR scores gradually decrease as $\alpha$ increases, and are both higher than the models trained without A2T. The best separation performance is achieved at an intermediate value of $\alpha$, e.g. $\alpha\in[0.3, 3]$, and a minor improvement on SNR and SI-SDR can be achieved at the best values of $\alpha$ across all overlap ratios. It confirms the ability of A2T to find ``better'' outputs on the equal-valued contours.

Table~\ref{tab:result-recognition} presents the WER on the 500-utterance test set. Based on the observation in Table~\ref{tab:result-separation}, we only report the SNR-A2T results with $\alpha=0.3, 1, 3$, as they all achieve on par or better separation performance than the standard SNR while have a much lower distortion on the direc-path signal. We observe that adding the A2T term can always leads to improved WER across all overlap ratios. Moreover, the overall WER increases as $\alpha$ increases, and the best overall performance is achieved when $\alpha=0.3$. Note that $\alpha=0.3$ and $\alpha=3$ both give the best separation performance in table~\ref{tab:result-separation}, and $\alpha=3$ leads to lower WER than $\alpha=0.3$ in two overlap ratio ranges. This indicates that different $\alpha$ for different overlap ratios might further improve the overall performance, however we leave this as a future work to verify. Nevertheless, the results confirm that with a similar level of SNR and SI-SDR, the actual WER can vary by as large as 16\% relatively ($[25, 50)\%$ overlap ratio), and further emphasize the importance of the constraint on the equal-valued contours.

The definition of direct-path signal can vary in different literatures. Defining the direct-path RIR filter as $\pm6$~ms of the first peak in the RIR filter is the same as \cite{qian2018deep}, however the range can be relaxed to cover more early reverberation components similar to \cite{drude2019sms}. Here we also investigate the effect of different definitions of direct-path signals. Table~\ref{tab:result-separation-20} shows the separation results on the same datasets as above, while the direct-path RIR filter is defined as $\pm20$~ms of the first peak in the RIR filter. Interestingly, the separation performance measured by SNR and SI-SDR are both worse than those in table~\ref{tab:result-separation}, and the autoencoding performance measured by TSNR and TSI-SDR, although on a different definition of direct-path signal, are also worse. The reason might be that autoencoding on the $\pm20$~ms direct-path signal also suffers the equal-valued contour problem, as the early reverberations may also cause minor distortions on the overall reconstruction. The recognition performance presented in Table~\ref{tab:result-recognition-20} also show that the WERs on $\pm20$~ms A2T are in general worse than those on $\pm6$~ms A2T, while still better than the standard SNR objective. Moreover, unlike the observation in $\pm6$~ms A2T that a larger $\alpha$ leads to worse WER, a larger $\alpha$ here leads to better performance. The reason behind this observation is yet to be revealed. To summarize, A2T prefers a more aggressive definition of direct-path signal.

\section{Conclusion}
\label{sec:conclusion}
In this paper, we investigated the ``equal-valued contours'' in commonly-used objectives and metrics for end-to-end reverberant speech separation, and proposed auxiliary autoencoding training (A2T) to control the distortion introduced to the direct-path signal in the targets. Equal-valued contours in commonly-used metrics such as SNR and SI-SDR can let the model choose undesired outputs, which might hurt both the separation performance and the generalization ability. A2T assumed that the separation was done by a linear operation on the mixture signal, and added an extra loss term during training to perform autoencoding on the direct-path signal of the targets. We evaluated single-channel separation models trained with and without A2T on both signal quality metric and speech recognition accuracy, and showed that when A2T was properly added, both the separation performance and speech recognition accuracy can be improved. We further investigated the effect of different definitions of ``direct-path signal'', and observed that a more aggressive definition led to better overall performance.

For future works, we would like to extend A2T into more architectures, e.g. end-to-end multi-channel separation models \cite{luo2020end, gu2020enhancing} and beamforming-based hybrid models \cite{ochiai2020beam, chen2020continuous}. Exploring alternative ways to add the constraints on the equal-valued contours is also an interesting topic. Moreover, the observation in the experiment results show that the direct-path preservation ability varies in different overlap conditions, thus it would also be interesting to see if a overlap-sensitive $\alpha$ can further improve the performance.


\bibliographystyle{IEEEtran}
\bibliography{refs}

\end{document}